\documentclass[12pt]{iopart}
\pdfoutput=1

\usepackage{iopams}
\usepackage{graphicx,xcolor}
\usepackage{booktabs}
\usepackage{microtype}
\usepackage{subfig}
\usepackage{lineno}

\begin{document}

\title[An exactly solvable coarse-grained model for species diversity]{An exactly solvable coarse-grained model for species diversity}

\author{Samir Suweis$^{1}$, Andrea Rinaldo$^{2,3}$ and Amos Maritan$^{1,*}$}

\address{$^1$ Dipartimento di Fisica G. Galilei, INFN and CNISM Universit\`a di
Padova, Via Marzolo 8  I-35151 Padova  Italy$^2$ Laboratory of Ecohydrology ECHO/ISTE/ENAC, Facult\'e ENAC,\'Ecole Polytechnique F\'ed\'erale Lausanne (EPFL) Lausanne (CH), $^3$ Dipartimento IMAGE Universit\`a di Padova I-35131 Padova Italy,$^*$ corresponding author: \it{maritan@pd.infn.it}}

\begin{abstract}
We present novel analytical results about ecosystem species diversity that stem from a proposed coarse grained neutral model based on birth-death processes. The relevance of the problem lies in the urgency for understanding and synthesizing both theoretical results of ecological neutral theory and empirical evidence on species diversity preservation. Neutral model of biodiversity deals with ecosystems in the same trophic level where per-capita vital rates are assumed to be species-independent. Close-form analytical solutions for neutral theory are obtained within a coarse-grained model, where the only input is the species persistence time distribution. Our results pertain: the probability distribution function of the number of species in the ecosystem both in transient and stationary states; the $n$-points connected time correlation function; and the survival probability, defined as the distribution of time-spans to local extinction for a species randomly sampled from the community. Analytical predictions are also tested on empirical data from a estuarine fish ecosystem. We find that emerging properties of the ecosystem are very robust and do not depend on specific details of the model, with implications on biodiversity and conservation biology.
\end{abstract}


\section{Introduction}
Statistical physics is decisively contributing to our understanding of ecological processes. In fact it is providing powerful theoretical tools and innovative steps towards the comprehension and the synthesis of broad empirical evidences on macro-ecological laws and emerging biodiversity patterns \cite{Chave2002,Holts2006,Dewar2008,zillio2008,McGill2010}. One field where statistical physics has been particularly successful is the neutral theory of biodiversity  \cite{UNTB,Bell2001,volkov2003,Azaele2006,volkov2007,Vallade2003,Houchmandzadeh2003,Houchmandzadeh2008}. This theory is based on the assumption that, within the same trophic level, per--capita vital rates are species-independent. It offers a unified theoretical framework for ecosystems dynamics by invoking solely basic ecological processes such as birth, death, migration and dispersal limitation \cite{UNTB,Bell2001}. Exact solutions have been found for several ecologically relevant quantities, such as: the relative species abundance distribution (RSA) \cite{Azaele2006,volkov2007,Mckane2004}; species spatial patterns and clustering \cite{Houchmandzadeh2003,Houchmandzadeh2008}; patterns of  $\beta$-diversity (i.e. intra- and inter-species spatial correlation) \cite{condit2002,Zillio2005}; species area-relationship (SAR) \cite{zillio2008,He2011}; and species' persistence time distributions  \cite{pigolotti2005,bertuzzo2011,suweis2012}. These results enabled the theory to be tested contrasting empirical data, and to assess the power of neutral models in predicting ecological patterns in many ecosystems.

Let consider a given trophic level (i.e. plants) in a local ecosystem. "Local" means a community immersed in a larger one, called meta-community, that is considered as infinite in size with respect to the local community (serving as a �reservoir�). We assume that each individual in the local community has a natural death rate $d$ and birth rate $b$  (linear birth and death rates). Moreover, due to speciation or diversification events (i.e. immigration from an outside region) �new� species (i.e. species not already present in the local community) enter in the system with rate $\lambda$.  

A general formulation of the neutral theory employs the birth-death master equation (ME) for the abundance dynamics of a species in the ecosystem \cite{volkov2003,Vallade2003,gardiner2004}.
  Several exact results are known for birth-death models where the population, $N$, of the local community is strictly conserved and/or the corresponding $\lambda$ scales as $N$ (e.g see the infinite alleles model with mutation \cite{Ewens2004} or the voter model \cite{Liggett1975,Durrett2002,McKane2007}).

In particular, a very well known and studied pattern in ecology is the RSA of the local community, defined as the fraction of species of a given abundance. It describes key elements of biodiversity, as the frequency of rare species in the ecosystem \cite{UNTB,Bell2001}. Another important quantity in conservation ecology is species richnesses, the number of different species in the community, independently of their abundance. Analytical solutions for both of these quantities can be found in the literature \cite{volkov2003,Vallade2003,Ewens2004,Kimura1970}.

Unfortunately, transient solutions for species richness and its $n$ points time correlations functions (i.e. the probability of having $S_1$ species at time $t_1$, $S_2$ species at time $t_2$, etc..) are not easy to calculate using this approach. Moreover, higher complications comes into play if we want to generalize the dynamics with non linear birth and death rates, i.e. birth and death rates which are not proportional to the population size \cite{Volkov2005}. For this reason, in the next section we introduce a coarse grained version of this model which will allow us to deduce in a simpler way novel analytical results about ecosystem biodiversity, where the only input is the species persistence time distribution. We note that, given the assumption of no interaction among species, no approximation is made to obtain these results, but they are limited to knowing whether a species is present/absent, as opposed to knowing how
many individuals there are of that species at a given time. In particular we obtain close-form analytical results for: the probability distribution function (pdf) of the number of species in the ecosystem in transient states; the $n-$point connected time correlation function; and the survival probability, $p_s(\tau)$, defined as the pdf of time-spans $\tau$ to local extinction for a species randomly sampled among the observed assemblages at a certain given time. A comparison of analytical predictions with empirical data from a estuarine fishes ecosystem will be carried out. A set of conclusions closes then the paper.

\section{Theoretical Framework}

\subsection{Grand-Canonical Formulation of Neutral Theory}
In the natural/realistic case where the total number of species ($S$) and individuals ($N$) in the local community is not fixed, we can described the abundance dynamics of the species in the system employing a "grand-canonical" formulation of the neutral theory. In particular, if $\phi_j$ gives the number of species with $j$ individuals, and 
$\vec{\phi}=\{\phi_1,\phi_2,...,\phi_{\infty}\}$, then the probability $P(\vec{\phi},t)$ - of having at time $t$, $\phi_1$ species with one individuals, $\phi_2$ species with two individuals, and so on - is univocally described
once the initial condition $P(\vec{\phi},t=0)$ is known and the transition rates from $\vec{\phi'}$ to $\vec{\phi}$ due to birth/death or speciation events are given. In particular, for $k\geq 1$ with birth rate $b_k=b\cdot k\cdot \phi_k$ (birth event within species of $k$ individuals) we have the transitions $\{\phi_k\rightarrow \phi_k-1; \phi_{k+1}\rightarrow \phi_{k+1}+1\}$, while with death rate $d_k=d\cdot k\cdot \phi_k$ (death of an individual belonging to species with $k$ individuals) we have the transitions $\{\phi_k\rightarrow \phi_k-1; \phi_{k-1}\rightarrow \phi_{k-1}+1\}$. Finally with rate $\lambda$ the transition $\{\phi_1\rightarrow \phi_1+1 \}$ occurs. We remark that in our framework, the total number of individuals $N$ is not fixed and the rate at which new species enters in the system is independent of $N$. 

In our grand-canonical formulation, the stationary solution of the  master equation corresponding to $\mathcal{P}(\vec{\phi},t)$ can be written as
$
\mathcal{P}_{staz}(\vec{\phi})=\frac{1}{\mathcal{Z}}\prod_{k}\mathcal{P}(\phi_k),
$
with
\begin{equation}\label{RSA}
   \mathcal{P}(\phi_k)= \left(\frac{\gamma x^k}{k}\right)^{\phi_k}\frac{1}{\phi_k!}; \qquad\quad \mathcal{Z}=e^{-\gamma\ln(1-x))},
\end{equation}
where $\gamma=\lambda/b$ and $x=b/d$. Note that $x$ represents the ratio of
effective per capita birth and death rate and if $\lambda \neq 0$ it has to be less than 1 in order to avoid demographic explosion. On the other hand if  $\lambda=0$, then at equilibrium
there are no individuals in the community because all species eventually go extinct \cite{volkov2003,Azaele2006}. 

In our theoretical framework the RSA is deduced by the first moments of $\mathcal{P}_{staz}$, while species richness is described by the probability of having  $s$ different species in the community at stationarity, that turn out to be Poisson distributed with mean $\langle S \rangle = - \gamma\ln(1-x)$:
\begin{equation}\label{pS_neutral}
   \mathcal{P}(s)= \sum_{\phi_1,\phi_2,...,\phi_{\infty}}\mathcal{P}(\vec{\phi})\delta_K(\sum_j \phi_j-s)=\frac{\left(-\gamma \ln(1-x)\right)^s}{s!}(1-x)^\gamma 
\end{equation}
where $\delta_K$ is the Kronecker delta, which is $1$ when its argument zero and zero otherwise.

Since to maintain a finite local community size, $x$ must be less than one, then each species is eventually doomed to extinction. We thus can define the persistence time $\tau$ of a species as the time incurred between its emergence in the system (due to diversification or immigration) and its extinction \cite{pigolotti2005,bertuzzo2011,suweis2012}. In particular, persistence times for a species that undergoes the linear birth-death dynamics presented above, are distributed according to the species persistence time (SPT) pdf \cite{pigolotti2005}, 
\begin{equation}\label{SPTpigolotti}
    p_{spt}(t)=d\left[\frac{1-x}{e^{d(1-x)t}-1}\right]^2 e^{d(1-x)t}.
\end{equation}
From now on, without loss of generality, we set $d=1$. From Eq. (\ref{SPTpigolotti}) we can express the mean persistence time, a key ecological quantity also known as mean extinction time \cite{pigolotti2005,Ewens2004,Kimura1970}, as $\langle \tau \rangle=\int_0^{\infty}\tau p_{spt}(\tau)\mathrm{d}\tau=-\ln(1-x)/x$. Finally, as expected,
the mean number of species at stationarity is related to $\langle \tau \rangle$, through Eqs. (\ref{pS_neutral}) and (\ref{SPTpigolotti}), by the simple formula $\langle S \rangle=\lambda\langle \tau \rangle$.  

\subsection{The Coarse-Grained Model}
A coarse-grained view of the grand-canonical formulation of the neutral theory can be described as follows. Each species $i$ within the local community emerges at a random time $\{t_i\}_{i\geq0}$ with a rate $\lambda$, that is, the probability a new species emerges in the infinitesimal time interval $\mathrm{d}t$ is $\lambda\mathrm{d}t$. Therefore the probability of having $k$ new species in the ecosystem up to time $t$, $U_{k}(t)$, is given by (see Appendix A):
\begin{equation}\label{nu}
    U_k(t)=e^{-\lambda t}\sum_{s_0=1}^{k}\frac{(\lambda t)^{k-s_0}}{(k-s_0)!},
\end{equation}
where $s_0$ is the number species at time $t=0$, i.e $U_k(t=0)=\delta_K(k-s_0)$. Then the newcomer species $i$ persists for a random duration $\tau_i$ where the random variables $\tau_i$ are independent and identically distributed with a given pdf $p_{spt}(\tau)$. Therefore species arrive as a Poisson process (with rate $\lambda$), and then depart after some non-Poissonian waiting time $\tau\sim p_{spt}$. 

\begin{figure}[h!]
\centering
\includegraphics[width=.98\columnwidth]{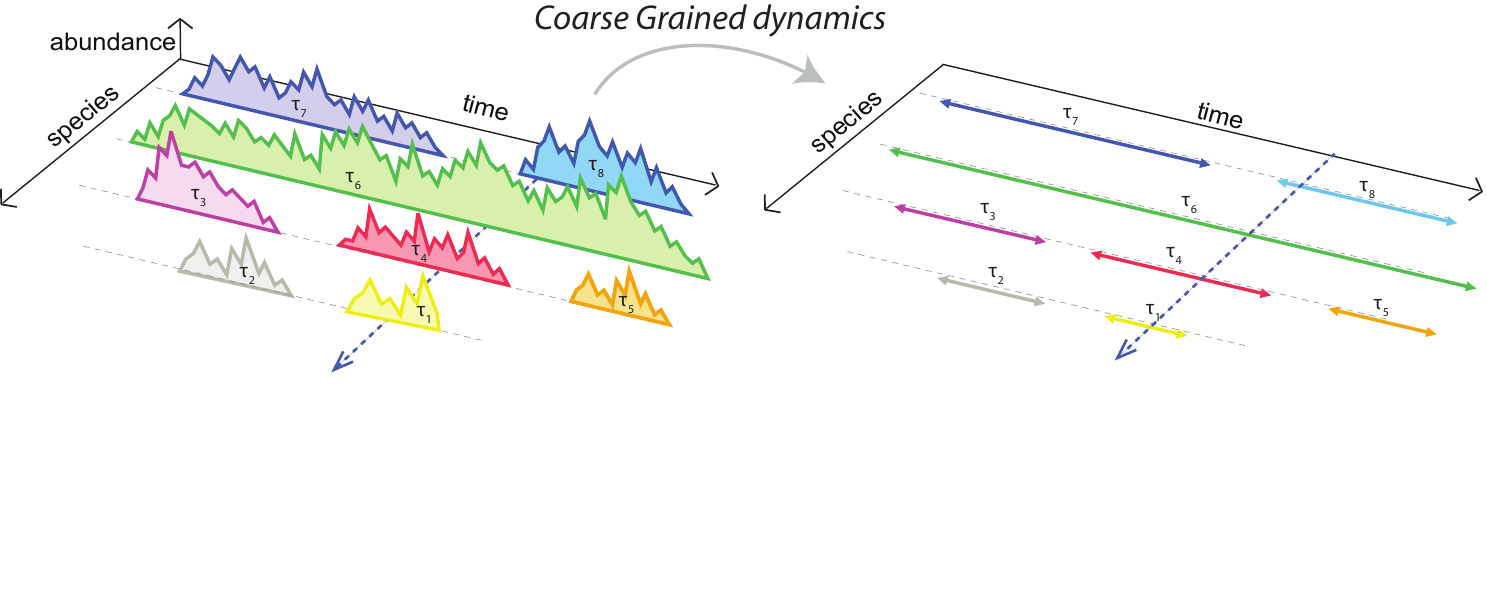}
\caption{a) Coarse grained view of ecosystem species dynamics. Species emerge in the ecosystem uniformly in time, and each persists for a random time $\tau$ drawn from the SPT pdf $p_{spt}$.} \label{fig1}
\end{figure}

This model is coarse grained in the sense that we do not take explicitly into account information on species abundance, implicitly contained in the functional form of $p_{spt}(\tau)$  (see Figure \ref{fig1}).  This approach enable us to go beyond Eq. (\ref{pS_neutral}), and to generalize these results also for general birth and death rates, by taking different functional shape of the SPT pdf. We remark that the mapping is exact only with respect to the grand-canonical formulation of the neutral theory, while, as we will show, is only a good approximation of the classic birth-death ME approach in the limit of large local community ($N$ fixed and $N\rightarrow\infty$).

\begin{figure}[htbp]
\begin{center}
\includegraphics[width=.98\columnwidth]{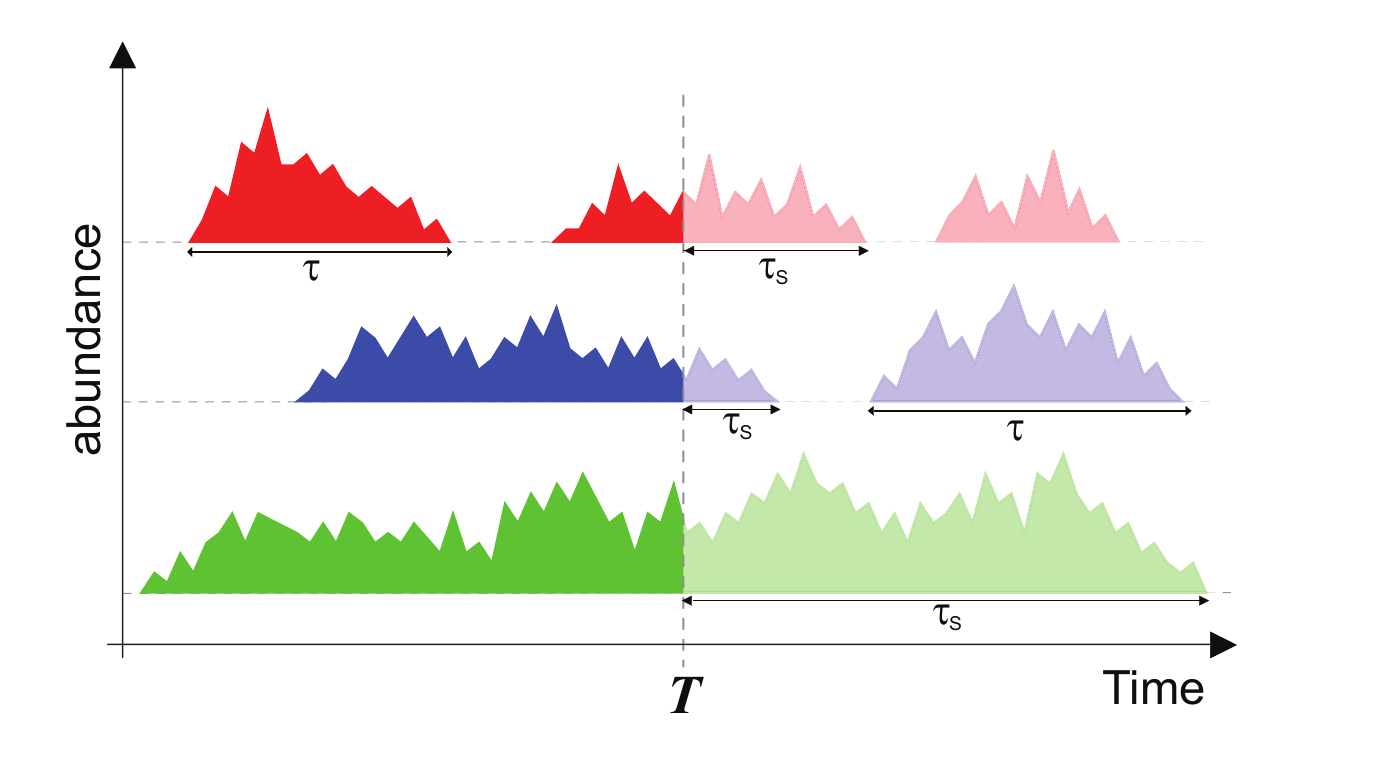}
\caption{Schematic representation of survival times $\tau_s$, defined as the time to local extinction
of a species randomly sampled among the observed assemblages at a certain time $T$. $\tau$, instead, denotes the persistence time of a species, and it is defined
as the time incurred between its emergence in the system and its extinction. Eq. (\ref{surv_final}) gives the relation between the two distributions, independently of the functional form of the birth and death rates.}
\label{fig2}
\end{center}
\end{figure}

Within our framework, the number of persistent species in the ecosystem at a given time $t$ (see Figure \ref{fig2}) is given by:
\begin{equation}\label{s}
    S(t)=\sum_{i=1}^{k(t)}\Theta(t_i+\tau_i-t),
\end{equation}
where $k(t)$ is the number of new species that entered into the system in the time interval $[0,t)$, i.e. $t_i<t$ for $i\leq k(t)$ and $t_i\geq t$ for $i> k(t)$. $\Theta(z)$ is the Heaviside step function, which is $1$ when its argument is positive and zero otherwise. $t_i$ is the time when the $i$-th species enters into the ecosystem and $t_i+\tau_i$ when it extincts. The probability of having $s$ species present in the ecosystem at time $t$ is thus $ \mathcal{P}(s,t)=\langle \delta_K(s-S(t))\rangle$, i.e.
\begin{equation}\label{p(S,t)}
 \mathcal{P}(s,t)=\sum_{k=0}^{+\infty} U_k(t)\int_0^t\prod_{i=1}^k
\frac{dt_i}{t}\int_0^{\infty}\prod_{j=1}^k d\tau_jp_{spt}(\tau_j)\delta_K\left(s-\sum_{i=1}^k\Theta(t_i+\tau_i-t)\right).\qquad\quad
\end{equation}
It is useful and customary to define the generating function of the process (the discrete Laplace transform), $\hat{\mathcal{P}}(z,t)=\sum_{s=0}^{\infty} z^s \mathcal{P}(s,t)$, and analogously for $U_k(t)$. Eqs. (\ref{nu}) and (\ref{p(S,t)}) lead to (see Appendix A)
\begin{equation}\label{log_gen_fun4}
    \hat{\mathcal{P}}(z,t)=\hat{U}\big(1+\frac{z-1}{t}f(t),0\big) e^{\lambda f(t)(z-1)},
\end{equation}
where
\begin{equation}\label{f(t)}
   f(t)=\int_{0}^{t}d\tau P_{>}(\tau)=\int_0^{+\infty}p_{spt}(\tau)\min[t,\tau]d\tau,
\end{equation}
with $P_{>}(t)=\int_t^{+\infty}p_{spt}(\tau)$. If, for example, we assume that $U_s(0)=\delta_K(s-1)$ (and therefore $\hat{U}(z,0)=z$), then from Eq. (\ref{log_gen_fun4}) we get
\begin{equation}\label{p(S,t)1}
    \mathcal{P}(s,t)=e^{-\lambda f(t)}\frac{(\lambda f(t))^s}{s!}\bigg(1+\frac{s-\lambda f(t)}{\lambda t}\bigg).
\end{equation}
One can see that $f(t)/t=\int_0^{\infty}p_{spt}(\tau)\hbox{min}(1,\tau/t)d\tau\rightarrow0$ in the $t\rightarrow \infty$ limit. This follows from $\hbox{min}(1,\tau/t)\leq1$, $\lim_{t\rightarrow\infty}\hbox{min}(1,\tau/t)=0$ for all $\tau$ and the Lebesgue's dominated convergence theorem (see for example \cite{Bartle1995}). Since $\hat{U}(z,0)$ is continuous in the interval $-1\leq z\leq1$ we have $\hat{U}((1+(z-1)/t)f(t),0)\rightarrow \hat{U}(1,0)=1$ in the large time limit, implying that the initial condition is forgotten in this limit, as expected. If $\int_0^{\infty}p_{spt}(\tau)\tau d\tau=\langle \tau \rangle<\infty$, then $\lim_{t\rightarrow\infty}f(t)=\langle \tau \rangle$ and from Eq. (\ref{log_gen_fun4}) we get $ \hat{\mathcal{P}}(z)=\lim_{t\rightarrow\infty}\hat{\mathcal{P}}(z,t)=e^{(1-z)\lambda\langle \tau \rangle}$, or equivalently
\begin{equation}\label{Pstaz}
    \mathcal{P}(s)=\lim_{t\rightarrow\infty}\mathcal{P}(s,t)=\frac{(\lambda \langle \tau\rangle)^s}{s!}e^{-\lambda \langle \tau\rangle},
\end{equation}
i.e. a Poisson pdf with mean given by the product between the species emergence rate $\lambda$ and the mean average persistence time. As expected, we found the same result given by Eq. (\ref{pS_neutral}).

From Eq. (\ref{log_gen_fun4}) we can easily calculate the average numbers of species at a generic time $t$. Indeed, if $\langle S \rangle_{t=0}=\sum_s s U_s(t=0)<\infty$, we have that $\hat{U}(z,0)$ admits left derivative at $z=1$ and taking the left derivative at $z=1$ of Eq. (\ref{log_gen_fun4}) we obtain
\begin{equation}\label{mean_S}
  \langle S \rangle_t=\frac{\partial}{\partial z}\hat{\mathcal{P}}(z,t)_{|_{z=1}}=\langle S \rangle_{t=0} \frac{f(t)}{t}+\lambda f(t)\stackrel{t\rightarrow\infty}{\rightarrow}\lambda \langle \tau \rangle,
\end{equation}
where again we see that the initial condition is forgotten in the large time limit. These results have been tested numerically (see Figure \ref{fig3}).

\section{Generating Function Approach}
 We are interested in the $n-$point connected time correlation function of our process (or cumulant, see \cite{Kubo1962} ) $\langle S(t^1)S(t^2)\cdot\cdot\cdot S(t^n)\rangle_C$ in the stationary conditions (min$_{i=1,...,n}t_i\rightarrow\infty$) and $t^{i+1}-t^i$ fixed, for $i=1,...,n-1$. The generating function is given by
 \begin{eqnarray}\label{GF-var}
   \nonumber \mathcal{Z}_T(\{h\})&=&\big\langle e^{-\int_0^{T}dt h(t)S(t)}\big\rangle=\sum_{k=0}^{+\infty} U_k(T)\int_0^{T}\prod_{i=1}^{k}\frac{dt_i}{T}\int_0^{+\infty}\prod_{j=1}^{k}d\tau_jp_{spt}(\tau_j)\times\qquad\\
    \nonumber&\times& \exp\{-\sum_{i=1}^k\int_0^{T}dt' h(t')\Theta(t_i+\tau_i-t')\Theta(t'-t_i)\}=\\
    &=&\,\sum_{k=0}^{+\infty} U_k(T)z(T)^k=\hat{U}\big(z(T),0\big)e^{\lambda T (z(T)-1)},
\end{eqnarray}
 where
 \begin{equation}\label{zT}
    z(T)=\int_0^{T}\frac{dt}{T}\bigg\langle \exp\bigg\{-\int_0^{T}dt' h(t')\Theta(t+\tau-t')\Theta(t'-t)\bigg\}\bigg\rangle_{\tau},
\end{equation}
 with $\langle \cdot \rangle_\tau=\int_0^{\infty}d\tau  \cdot p_{spt}(\tau)$.

 The $n-$point correlation function is obtained by choosing
\begin{equation}\label{h_t}
    h(t)=\sum_{i=1}^n\delta(t-t^i)h_i\qquad\hbox{with}\;h_i\geq0
\end{equation}
 and
 \begin{equation}\label{npoint}
\lim_{t^1\rightarrow\infty}\langle S(t^1)S(t^2)\cdots S(t^n)\rangle_C=\lim_{T\rightarrow\infty}\frac{\partial^n}{\partial h_1\cdots \partial h_n} \ln\big[\mathcal{Z}(\{h\})\big]_{|_{z=0}},
\end{equation}
 where we assume $0<t^1\leq t^2 \leq \cdots \leq t^n<T$.

We will show that in the stationary conditions $i)$ $\lim_{T\rightarrow \infty}z(T)=1$ and $ii)$ if $\langle\tau\rangle_{\tau}<\infty$,
\begin{equation}\label{npoint2}
   \langle S(t^1)S(t^2)\cdot\cdot\cdot S(t^n)\rangle_C=\lim_{T\rightarrow\infty}\lambda T \frac{\partial^n}{\partial h_1\cdot\cdot\cdot \partial h_n} z(T)_{|_{h_i=0}}.
\end{equation}
Using Eqs. (\ref{zT}) and (\ref{h_t}) we obtain
\begin{eqnarray}\label{zT2}
  \nonumber z(T) &=& \int_0^T\frac{dt}{T}\big\langle\exp\big\{-\sum_{i=1}^nh_i\Theta(t+\tau-t^i)\Theta(t^i-t)\big\}\big\rangle_{\tau}\\
 \label{zT3} &=&\int_0^T\frac{dt}{T}\bigg\langle\prod_{i=1}^n\bigg[\big(e^{-h_i}-1\big)\Theta(t+\tau-t^i)\Theta(t^i-t)+1\bigg]\bigg\rangle\\
 \label{zT4} &=&1 +\sum_{k=1}^{n}\sum_{i_1<i_2<\cdots<i_k}\bigg\langle \hbox{max}\big\{0,\frac{t^{i_1}-\hbox{max}\{0,t^{i_k}-\tau\}}{T}\big\}\bigg\rangle_{\tau}\prod_{j=1}^k(e^{-h_{i_j}}-1),\qquad
\end{eqnarray}
where we have expanded the product on the r.h.s. of Eq. (\ref{zT3}) as $\prod_{i=1}^n(1+v_i)=1+\sum_{i=1}^nx_i+\sum_{i_1<i_2}^nx_{i_1}v_{i_2}+\cdots=1+\sum_{k=1}^n\sum_{i_1<i_2<\cdots i_k}v_{i_1}v_{i_2}\cdots v_{i_k}$. Using the Lebesgue's dominated convergence theorem, the average in Eq. (\ref{zT4}) tends to zero in the $T\rightarrow \infty$ limit and thus $\lim_{T\rightarrow\infty}z(T)=1$, which proves $i)$. Let us consider now
\begin{equation}\label{zT5}
\lim_{T\rightarrow\infty}T(z(T)-1)=\lim_{T\rightarrow\infty}\sum_{k=1}^{n}\sum_{i_1<i_2<\cdots<i_k}\bigg\langle \hbox{max}\big\{0,t^{i_1}-\hbox{max}\{0,t^{i_k}-\tau\}\big\}\bigg\rangle_{\tau}\prod_{j=1}^k(e^{-h_{i_j}}-1).\qquad
\end{equation}
Since $\hbox{max}\big\{0,t^{i_1}-\hbox{max}\{0,t^{i_k}-\tau\}\big\}<\tau$ and $\lim_{t^1\rightarrow\infty}\hbox{max}\big\{0,t^{i_1}-\hbox{max}\{0,t^{i_k}-\tau\}\big\}=\hbox{max}\{0,\tau-(t^{i_k}-t^{i_1})\}$, again for the Lebesgue dominated convergence theorem, Eq. (\ref{zT5}) leads to
\begin{equation}\label{zT6}
\lim_{T\rightarrow\infty}T(z(T)-1)=\sum_{k=1}^{n}\sum_{i_1<i_2<\cdots<i_k}\prod_{j=1}^k(e^{-h_{i_j}}-1)\big\langle\hbox{max}\{0,\tau-(t^{i_k}-t^{i_1})\}\big\rangle_{\tau}\equiv \mathfrak{F}(\{h\}),\quad\\
\end{equation}
and therefore we have that Eq. (\ref{GF-var}) becomes
\begin{equation}\label{staz_funz_gen}
    \lim_{T\rightarrow\infty}\mathcal{Z}_T(\{h\})=e^{\lambda \mathfrak{F}(\{h\})}.
\end{equation}
Using Eqs. (\ref{npoint}) and (\ref{staz_funz_gen}) we finally prove Eq. (\ref{npoint2}). In particular, we note that in the large time limit, the $n-$point connected time correlation function
  \begin{equation}\label{npoint3}
 \lim_{t^1\rightarrow\infty} \langle S(t^1)S(t^2)\cdot\cdot\cdot S(t^n)\rangle_C=\lambda \big\langle\big[\tau-(t^n-t^1)\big]\Theta(\tau-(t_n-t_1))\big\rangle_{\tau}
\end{equation}
is independent of $t^2\cdots t^{n-1} \in (t^1,t^n)$.

\section{Survival times pdf}
The survival time $\tau_s$ is defined as the time to local extinction of a species randomly sampled among the observed assemblage of species at a certain time $T$ (see Figure \ref{fig2}).

We can express  $\tau_s$ as a function of the random variables $t_0$ (emergence time of that species) and $\tau$ for which the pdf is known:
\begin{equation}\label{tau_s}
    \tau_s=t_0+\tau-T \;\;\;\; \hbox{if $0<t_0<T$ and $t_0+\tau\ge T$.}
\end{equation}
We then express the survival time (ST) pdf conditional to a persistence $\tau$ as:
$$
    p_{s}(t|\tau)=\mathcal{C}\langle \delta(t-(t_0+\tau-T))\Theta(t_0+\tau-T)\Theta(T-t_0)\Theta(t_0)\rangle,
$$
where the constant $\mathcal{C}$ ensures normalization. Solving the ensemble average operators yields:
$$
    p_{s}(t|\tau)=\mathcal{C}\Theta(\tau-t)\Theta(t-\tau+T)\Theta(t),
$$
and, by marginalizing over $\tau$, we obtain the ST pdf:
\begin{equation}\label{p_tau_s}
    p_{s}(t)=\frac{1}{\langle\hbox{min}(\tau,T) \rangle_{\tau}}\int_{t}^{t+T}p_{spt}(\tau)d\tau.
\end{equation}
Without loss of generality, it can be assumed that $T$ is not affected by the boundary condition in $t=0$, i.e. $T\rightarrow\infty$ and $p_{s}(t)=\frac{1}{\langle\tau \rangle_{\tau}}\int_{t}^{\infty}p_{spt}(\tau)d\tau$. Particularizing now to the case of persistence distributions for linear birth-death processes (Eq. (\ref{SPTpigolotti})), the survival pdf asymptotic behavior is:

\begin{equation}\label{survivor}
    p_{s}(t)\propto \int_{t}^{\infty}\tau^{-2}((1-x)\tau/(1-e^{-(1-x)\tau}))^2 e^{-(1-x)\tau}d\tau\propto\frac{1}{e^{(1-x) t}-1}\sim \left\{
                   \begin{array}{ll}
                     t^{-1}, & \hbox{for $t\ll t^*$} \\
                     e^{-t/t^*}, & \hbox{for $t\gg t^*$}
                   \end{array}
                 \right.
\end{equation}

where $t^*=(1-x)^{-1}$.

\section{Applications and Comparison with Empirical Data}

\subsection{Gamma Species Persistence Time Distribution.}
Recent results \cite{bertuzzo2011,suweis2012} have shown that SPT distribution for several different type of ecosystems exhibit power-law behavior with an exponential$-$like cut-off:
\begin{equation}\label{Lifetime_bb}
    p_{spt}(t)=\mathcal{A}^{-1}\, t^{-\alpha}e^{-(1-x) t}\theta(t-\tau_0),
\end{equation}
where $\mathcal{A}=(1-x) ^{\alpha -1} \Gamma (1-\alpha ,(1-x)\tau_0)$ is the normalization constant, and $\Gamma (a,b)$ is the incomplete Gamma function. The exponent $\alpha$ of the power law is suggested to depend on the spatial structure of the embedding ecosystem. SPT distributions  exhibit progressively smaller scaling exponents $\alpha$ for increasing constraints in the connectivity structure of the environmental matrix \cite{bertuzzo2011}. Specifically, numerical simulations \cite{bertuzzo2011} show that the exponents range between $\alpha=2$, that corresponds to the case of global dispersal (mean field), $\alpha=1.91 \pm 0.01$ for a 3D lattice, $\alpha=1.83 \pm 0.02$ in a savannah (2D lattice), $\alpha=1.64 \pm 0.02$ for river network topology, up to $\alpha=3/2$ for 1D systems.

For the functional shape of the SPT distribution given by Eq. (\ref{Lifetime_bb}) with $\alpha\leq2$, the following asymptotic limits are obtained:
\begin{eqnarray}\label{cum_life_lim1}
     P_{>}(t)&=&\int_t^{\infty}p_{spt}(t')dt'=\langle \tau \rangle_{\tau} p_{s}(t)=\\
     \nonumber&=&\left\{
                   \begin{array}{ll}
                     \frac{1}{\Gamma(1-\alpha,(1-x)\tau_0)}e^{-(1-x) t}((1-x) t)^{-\alpha}(1-\alpha \frac{1}{t(1-x)}+o\left(\frac{1}{t(1-x)}\right)^2, & \hbox{for $t\gg t^*$}\\
                     (\tau_0/t)^{\alpha-1}, & \hbox{for $\tau_0<t\ll t^*$}
                   \end{array}
                 \right.
\end{eqnarray}
while $P_>(t)\sim1$ for $t\rightarrow0$. The average persistence time follows from Eq.(\ref{f(t)})
\begin{equation}\label{f(t)_ex}
   \langle\tau\rangle = \int_0^{+\infty}d\tau P_{>}(\tau)=f_{\infty}=\frac{\tau _0 E_{\alpha -1}\left((1-x)  \tau _0\right)}{E_{\alpha }\left((1-x)  \tau
   _0\right)},
\end{equation}
where $E_{n}(z)=\int_{1}^{\infty}e^{-z t}/t^n dn$ is the exponential integral function. Setting $\tau_0=1$ for simplicity in Eq. (\ref{Lifetime_bb}), the corresponding two point correlation function derived from (\ref{npoint3}) is:
\begin{equation}\label{corr_ex1}
    \mathrm{cov}_S(t)=\frac{\lambda}{(1-x)} \frac{\Gamma (2-\alpha ,t (1-x) )-t (1-x)  \Gamma (1-\alpha ,t (1-x) )}{\Gamma (1-\alpha ,1-x )},\qquad \hbox{for }t>\tau_0.
\end{equation}
Therefore, the covariance is a decreasing function of $t$, that for large times goes to zero, i.e., $ \lim_{t\rightarrow+\infty} \mathrm{cov}_S(t)\sim e^{- (1-x) t} t^{-\alpha}=0.$

\subsection{Scale-Free SPT distributions}
An interesting case is obtained when $x\rightarrow 1$, i.e. the system is in the scaling regime where it does not exhibit a characteristic time scale. This is typically reported when SPT of families or genera -- as opposed to species -- are considered, possibly measuring them from the fossil record  \cite{newman1999}. Thus, longer time scales and long time$-$series must be considered.

Under such assumption, the normalized SPT pdf reads as:
\begin{equation}\label{scale_free}
    p_{spt}(t)=(\alpha-1)\,\tau_0^{\alpha-1} t^{-\alpha}\theta(t-\tau_0); \qquad\;\alpha>1,
\end{equation}
whereas the cumulative SPT distribution is
$
    P_{>}(t)=(\frac{\tau_0}{t})^{\alpha-1}\theta(t-\tau_0)+\theta(\tau_0-t),
$
and from Eq.(\ref{f(t)}) we obtain:
\begin{equation}\label{f(t)_ex2}
  f(t)=\left\{
         \begin{array}{ll}
           \min[\tau_0,t]+\tau_0\theta(t-\tau_0)\frac{(t/\tau_0)^{2-\alpha}-1}{2-\alpha}, & \hbox{$\alpha\neq2$;} \\
            \min[\tau_0,t]+\tau_0\theta(t-\tau_0)\ln(t/\tau_0), & \hbox{$\alpha=2$.}
         \end{array}
       \right.
\end{equation}
For this case, and in the limit $T\gg t$, the ST pdf given by Eq. (\ref{p_tau_s}) becomes
\begin{equation}\label{ptau_sPL}
  p_{s}(t)=\left\{
         \begin{array}{ll}
           \frac{2-\alpha}{T}\left(\frac{t}{T}\right)^{1-\alpha}, & \hbox{$\alpha\neq2$;} \\
           \frac{1}{t \ln(t/\tau_0)}, & \hbox{$\alpha=2$.}
         \end{array}
       \right.
\end{equation}

We note that two different cases can be pointed out. 1) $\mathcal{P}_{stat}(S)$ exists and depends on a microscopic time-scale $\tau_0$. This imply $\alpha>2$, leading to $ \lim_{t\rightarrow+\infty}f(t)=\tau_0\frac{\alpha-1}{\alpha-2}\equiv f_{\infty}$. 2) For $\alpha\leq2$ the stationary pdf $\mathcal{P}_{stat}(S)$ does not exists. In fact in this case $f(t)\stackrel{t \gg \tau_0}{\sim} \tau_0(\frac{t}{\tau_0})^{2-\alpha}\stackrel{t\rightarrow\infty}{\rightarrow}\infty$ for $\alpha<2$ and $ f(t)=\tau_0\ln\big(\frac{\max[t,\tau_0]}{\tau_0/e})\stackrel{t\rightarrow\infty}{\rightarrow}\infty$ ($e$ is the Neper number). The two point correlation function is thus obtained as:
\begin{equation}\label{corr_ex2}
    \mathrm{cov}_S(t)=\lambda (\alpha-1)\,\tau_0^{\alpha-1}\int_t^{+\infty}\tau^{-\alpha}(\tau-t)d\tau=\left\{
         \begin{array}{ll}
            \lambda\frac{\tau_0^{\alpha-1}}{\alpha-2}t^{2-\alpha}, & \hbox{$\alpha>2$;} \\
           \infty, & \hbox{$\alpha\leq 2$.}
         \end{array}
       \right.
\end{equation}
leading to    $\lim_{t\rightarrow+\infty} \mathrm{cov}_S(t)=\left\{
                        \begin{array}{ll}
                          0, & \hbox{$\alpha>2$;} \\
                          +\,\infty, & \hbox{$\alpha\leq2$.}
                        \end{array}
                      \right.$

\subsection{Hinkley Fish estuarine Database}
To test the results of the coarse grained neutral null model, we employ a long-term monthly database of a estuarine fishes ecosystem. Fish samples were collected from the cooling-water filter screens at Hinkley Point Power Station, located on the southern bank of the Bristol Channel in Somerset (England). The data span the period from January 1981 to January 2010. A full description of the intake configuration and sampling methodology  is given in \cite{Henderson2010}. A matrix $P$ is built using presence-absence records for each months during the 29 years. Each element $P_{st}$ of the matrix is equal to 1 if species $s$ is observed during month $t$, otherwise $P_{st}$ = 0. The empirical persistence time are defined as the number of consecutive months in which the measurements reveal the presence of the species. The presence-absence time series thus form a vector of length $T$, where $T$ is the total number of moths of observation.

\begin{figure}[htbp]
\begin{center}
\includegraphics[width=.98\columnwidth]{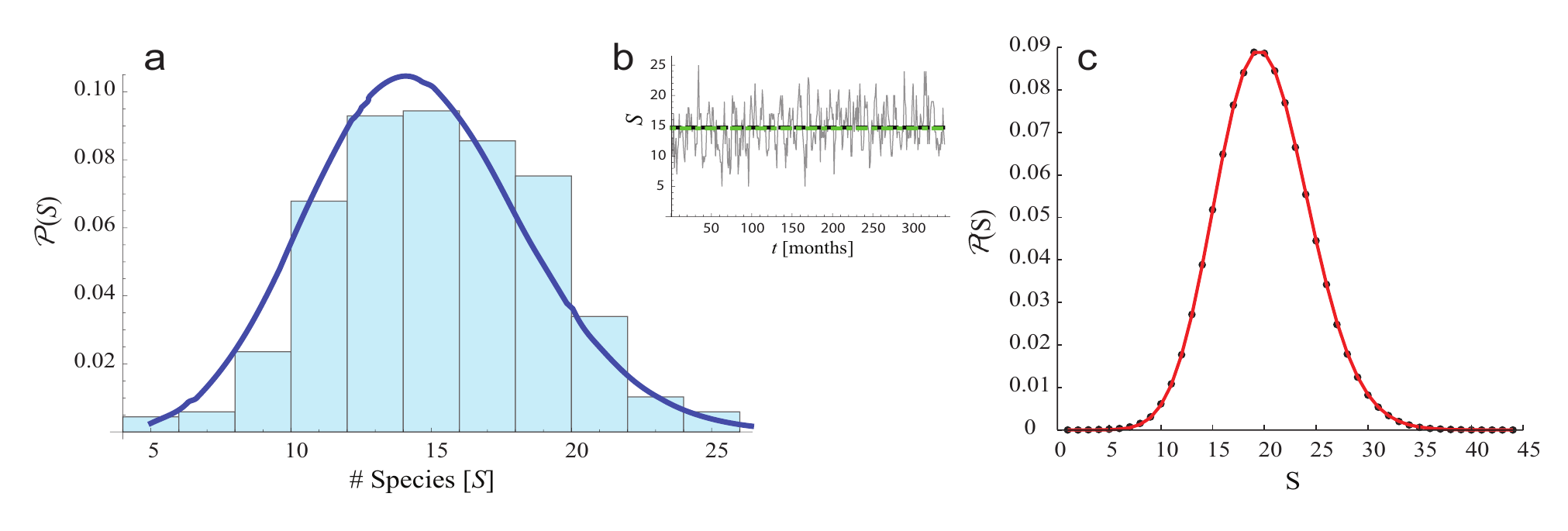}
\caption{a)Comparison between the empirical SPT pdf (histogram) and theoretical $\mathcal{P}(s)$  given by Eq. (\ref{Pstaz}) (blue line). b) Monthly time series of the analyzed estuarine fish ecosystem. The black dashed line represent the mean number of species $S_t$, while the green dot-dashed line the predicted average $\langle S \rangle$. c) The analytical solution  for  $\mathcal{P}(s)$ has been verified numerically.}
\label{fig3}
\end{center}
\end{figure}

\begin{figure}[h!]
\begin{center}
\includegraphics[width=.48\columnwidth]{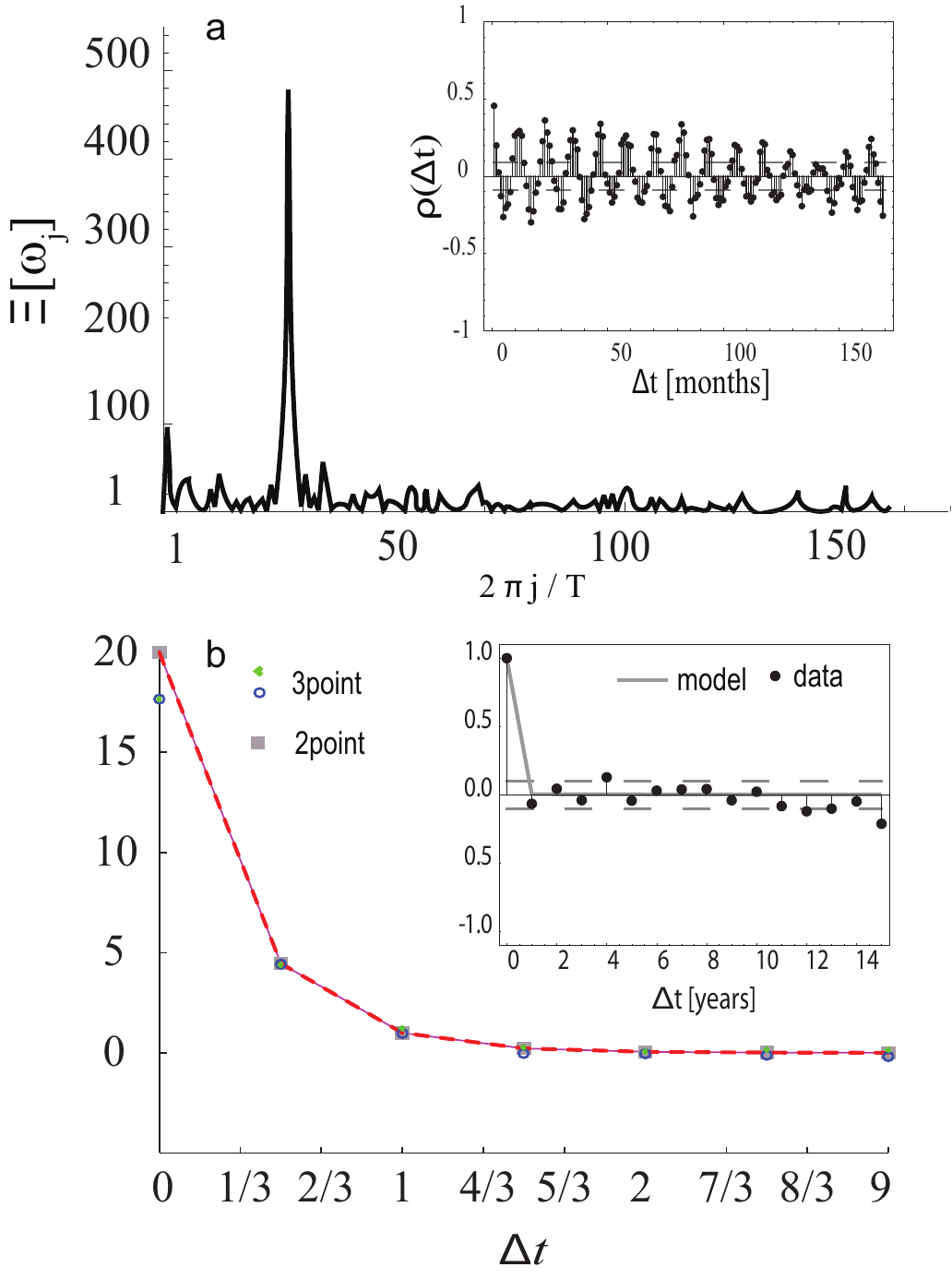}
\caption{a) Power spectrum ($\Xi[\omega_j]$) analysis of the time series $S_t$ and of the autocorrelation function $\rho(\Delta t)$ given by Eq. (\ref{autocorr}) (in the inset). We find a periodic behavior of the time-series  with period $t_p\approx 12$ months. b) We test the analytical result of the $n-$point correlation function (red dashed line) given by Eq. (\ref{npoint3}) for $p_{spt}(t)=\frac{1}{\tau_0}e^{-t/\tau_0}$ via numerical simulations. The parameters used in the simulation are   $T=1000$, $\tau_0=2$ and $\lambda=10$. Gray squares represents the two point  $\langle (S(t)-\bar{S})(S(t+\Delta t)-\bar{S})\rangle$ function, while green and blue dots are the three point correlation function  $\langle (S(t)-\bar{S})(S(t+\alpha\Delta t)-\bar{S})(S(t+\Delta t)-\bar{S})\rangle$, with $\alpha=1/3,2/3$, respectively. In the inset: comparison between empirical autocorrelation function $\bar{\rho}(\Delta t)$ (that does not take in account seasonal periodicity in $S_t$) (black dots) and analytical autocorrelation function predicted by the coarse grained model using Eq. (\ref{corr_ex1}) with $(1-x)=0.0001$, $\alpha=2$ and $\lambda=4.83$ (gray solid line). The dashed gray horizontal lines represent the 5 \%  confidence interval with respect to $\rho=0$.}
\label{fig4}
\end{center}
\end{figure}

Persistence time is defined as the length of a contiguous sequence of $1$s in the time series. From such time series we can thus reliably  estimate the empirical average persistence time $\langle \tau \rangle$ and the species emergence rate $\lambda$. For the analyzed fish estuarine ecosystem we find $\langle\tau\rangle=3.02$ [month] and $\lambda=4.83$ [month$^{-1}$]. Moreover, by summing  over rows of the matrix $P_{st}$, we obtain the total number of observed species on month $t$, i.e., $S_t=\sum_{s}P_{st}$ . We can thus calculate (assuming stationarity) the distribution of the number of persistent species in the system (Figure \ref{fig3}a), its first moments, $\bar{S}=\sum_{t=1}^TS_t/T$ (Figure \ref{fig3}b), and the empirical two-point correlation function (Figure \ref{fig4}):
\begin{equation}\label{autocorr}
  \rho(\Delta t)=\frac{\sum _{t=\Delta t+1}^TT \left(S_t-\bar{S} \right) \left(S_{t-\Delta t}-\bar{S}\right)}{(T-\Delta t) (\sigma^2_S)},
\end{equation}
where $\Delta t$ is the time lag and $\sigma^2_S=\frac{1}{T}\sum _{t=1}^T \left(S_t-\bar{S} \right)^2$, so that $\rho(0)=1$.

The average number of species predicted by the model can be obtained by Eq. (\ref{mean_S}), $\langle S \rangle=\lambda \langle\tau\rangle=14.6$, whereas the standard deviation is $\sqrt{\lambda \langle\tau\rangle}=\pm3.8$ that is in good agreement with the observed ecosystem diversity $\bar{S}\pm\sigma_S=14.7\pm3.7$ calculated from the  presence-absence matrix $P_{st}$ (see Figure \ref{fig3}a). The null hypothesis of a Poisson species distribution given by Eq. (\ref{Pstaz}) is accepted by both the Kolmogorov-Smirnov test and the $\chi^2$ test within a 95\% confidence interval ($P_{value}=0.05$). This analysis suggests that species diversity data cannot be used by themselves to discriminate among different mechanisms of demographic growths, i.e type of birth-death processes. Similar conclusion have been achieved using RSA data \cite{Volkov2005}.

The autocorrelation reveals a relevant periodic behavior in the species times series (Figure inset \ref{fig4}a). To investigate such periodicity we study the power spectrum of the time$-$series $\tilde{S}_t=S_t-\bar{S}_t$, i.e.  $\Xi[\omega_j]=\frac{1}{2\pi T}|\chi(\omega_j)|^2$, where $\chi(\omega_j)=\sum_{t=1}^T\tilde{S}_t e^{i \omega_j t}$ is the Fourier transform of $S_t$ and $\omega_j=2j\pi/T$ ($j=1,2,..T$) is the frequency. The spectrum exhibits a peak for $j\approx 30$, indicating that the period is $t_p\approx 12$ months (see Figure \ref{fig4}a).
Therefore the periodicity is a trivial effect due to seasonality of weather patterns and is not related to the fluctuations or intrinsic noise of the system dynamics \cite{boland2008}. To test Eq. (\ref{corr_ex1}) against empirical data, we then smooth out the periodicity due to seasonality. In order to do that, we calculate the empirical  autocorrelation function only considering a specific month for every year and then averaging over all twelve months, i.e., $\bar{\rho}^j(\Delta t)= \sum _{i=\Delta t+1}^{M_j} M_j \left(S^j_{i}- \bar{S}_j \right) \left(S^j_{i-\Delta t}-\bar{S}_j\right)/((M_j-\Delta t) (\sigma^2_j))$, and $\bar{\rho}=\sum_{j}\bar{\rho}_j/12$, where $j=\{Jan,Feb,...,Dec\}$ and $M_j$ is the total number of  $j-$type month in the whole time series. As can be seen from the inset in Figure \ref{fig4}b, once we remove seasonality Eq. (\ref{corr_ex1}) describes well the autocorrelation function $\bar{\rho}$.

An analysis of the fish SPT distribution have been also carried out. We find that SPT distributions are well described by the Gamma SPT with parameters compatible with $\alpha\approx 2.$ and  $x=0.9999$. This result holds with particular accuracy only when large SPT are considered. In fact, at monthly time scales, seasonality affects short SPTs, increasing the slope of the first part of the SPT pdf. Due to the negligible  effect of the cut-off at the monthly time scale, we find that also a power-law SPT  with $\alpha\approx 1.9$ fits the empirical SPT pdf. Note that here we are only interested in the estimation of the input parameters of the coarse grained neutral model, $\langle\tau\rangle$ and $\lambda$, as they can be calculated directly from the SPT measured time$-$series without the need of fitting the entire SPT distribution.

\section{Robustness of the results}
\subsection{Agreement with mean field voter model results.}
As we claimed in section 2, the proposed model is a coarse grained view of the grand-canonical version of neutral theory. Nevertheless we now show that our coarse grained model give a good approximation of results stemmed from the mean field approximation of the voter model with speciation  \cite{Zillio2005,Liggett1975,Durrett2002} or, equivalently the infinite alleles model with mutation \cite{Ewens2004} (for details on the mapping between these two models we refer to \cite{McKane2007}). The scheme of these models is the following. Consider a local community of $N$ individuals. At every time step a randomly selected individual in the ecosystem dies. With probability $1-\nu$ the individual is replaced by a colonizing offspring of another individual, randomly selected within the ecosystem whereas with probability $\nu$ the offspring belongs to a new species. For $N$ fixed and $N\rightarrow \infty$ the above dynamics is described a linear birth-death ME with $b=1-\nu$ and $d=1$ (and thus $x=b/d=1-\nu$):
\begin{equation}
\label{bdME}\frac{dp_n^{(i)}(t)}{dt}=x(n-1)p_{n-1}^{(i)}(t)+(n+1)p_{n+1}^{(i)}(t)-(x n+n)p_n^{(i)}(t)\quad \hbox{for\,}n\geq 1\qquad\\
\end{equation}
where $i=1,2,...,S\leq N$ and $p_n^{(i)}(t)$ is the probability for the $i^{th}$ species of having an abundance $n$ at time $t$. For the hypothesis of neutrality we have that $p_n^{(i)}(t)$ can be expressed as:
$$
    p_n^{(i)}(t)=p_n^*(t-t_i),
$$
where $p_n^*(t)$ is the probability for a single species to have abundance $n$ at time $t$ after its emergence (under the neutral assumption $p_n^*(t)$ is species invariant). The average population of the $i$ -th species, given that it is still present at time $t$, is $\langle n_i \rangle=\sum_{n\geq1}n p_n^{(i)}(t)/\sum_{m\geq1} p_m^{(i)}(t)$. The mean population size at time $t$ for the mean field voter model can be calculated as:
\begin{equation}\label{mean_n}
  \langle n(t) \rangle=\bigg\langle \frac{\sum_{i=1}^{D(t)}\sum_{n=0}^{+\infty}p_n^*(t-t_i)n}{\sum_{i=1}^{D(t)}\sum_{n=1}^{+\infty}p_n^*(t-t_i)}\bigg\rangle
\sim\frac{\big\langle\sum_{i=1}^{D(t)}\sum_{n=0}^{+\infty}p_n^*(t-t_i)n\big\rangle}{\big\langle\sum_{i=1}^{D(t)}\sum_{n=1}^{+\infty}p_n^*(t-t_i)\big\rangle},
\end{equation}
where $D(t)$ is the number of diversification events occurred until time $t$ and the ensemble average is over the random variables $t_i$ and $D(t)$. By using the fact that for all $i$, $t_i$ is Poisson distributed with the same frequency $\lambda$ and $U_D(t)$ is the pdf of the $D-$variable (see Eq. (\ref{nu})), then equation (\ref{mean_n}) simplifies to
$
  \langle n(t) \rangle= \int_0^tdt'\sum_{n}np_n^*(t')/\int_0^tdt'\sum_{n\geq1}p_n^*(t').
$ Thus, from the definition of  $p_n^*(t)$, it  follows that: $\sum_{n}np_n^*(t')=\langle n^* \rangle_t'$ or, the mean population of a species after a time $t$ from its emergence. Using Eq. (\ref{bdME}) it obeys the deterministic equation $  d\langle n^* \rangle_t/dt=-(1-x) \langle n^* \rangle_t$, which solution is $\langle n^* \rangle_t=\langle n^* \rangle_0\exp(-(1-x) t)$ \cite{pigolotti2005} . Thus we have that $\int_0^t\langle n^* \rangle_0\exp(-(1-x) t')dt'=\langle n^* \rangle_0\;(1-\exp(-(1-x) t))/(1-x)$. We observe that $\sum_{n\geq1}p_n^*(t')$ is the probability that the species has more than one individual at time $t$, that is the cumulative distribution of the SPT pdf $P_{>}(t)=\sum_{n\geq1}p_n^*(t)=\int_t^{+\infty}p_{spt}(\tau)d\tau$ and therefore
\begin{equation}\label{trick}
\int_0^tdt'\sum_{n\geq1}p_n^*(t')=\int_0^t dt' \int_{t'}^{+\infty}p_{spt}(\tau)d\tau=\langle \hbox{min}(\tau,t)\rangle_{\tau}=f(t).
\end{equation}
Using the above relations and $\langle n^* \rangle_0=1$ (since by definition $p_n^*(t=0)=\delta_K(n-1)$), we obtain
\begin{equation}\label{nmedio}
  \langle n(t) \rangle=\frac{1}{(1-x) f(t)}\stackrel{t\rightarrow+\infty}{\rightarrow}\frac{1}{(1-x)\langle\tau\rangle},
\end{equation}
 from which it follows that $\langle S \rangle=N/\langle n \rangle=N (1-x)\langle\tau\rangle$. Therefore if we approximate $N$ with the average number of individuals in the corresponding grand-canonical ensamble Eq. (\ref{pS_neutral}), i.e. $\langle N \rangle=x \partial_x \ln\mathcal{Z}=\lambda/(1-x)$, we found that Eq. (\ref{nmedio}) is the same result we have obtained from the coarse grained neutral model (compare with Eq. (\ref{mean_S}).


Interestingly, it has also been found that, for linear birth and death processes, the survival probability function has the same asymptotic functional shape given by Eq. (\ref{survivor}) \cite{pigolotti2005}. In fact, by assuming an initial population distribution given by the Fisher log-series $\mathcal{P}_{RSA}(n)=-x^n/(n \log(1-x))$ \cite{volkov2003}, and defining $p_{spt}(t\mid n_0)$  as the probability that a species starting with $n_0$ individuals is still present at time $t$, it has been shown that \cite{pigolotti2005}:
\begin{equation}\label{surivor_mean_field}
   p_{s}(t)=\sum_{n_0=1}^{\infty}p_{spt}(t\mid n_0)\mathcal{P}_{RSA}(n_0)=\left\{
                                                                                 \begin{array}{ll}
                                                                                   t^{-1}, & \hbox{for $t\ll t^*$;} \\
                                                                                   e^{-(1-x) t}, & \hbox{for $t\gg t^*$,}
                                                                                 \end{array}
                                                                               \right.
\end{equation}
that is, it has the same asymptotic behavior of the ST pdf as obtained in Eq. (\ref{survivor}) from the coarse grained model.

\subsection{Universal relation between persistence and survival distributions}

The fact that the survival probability function has the same asymptotic behavior in two different neutral models suggests that, rather than by chance, it possibly happens as a consequence of a deeper, and more general relationship between the SPT distribution and the ST probability function, valid regardless of the specific birth and death processes assumptions or, in other words, independently of the functional shape of $b(n)$ and $d(n)$.

To address this issue, we start by calculating the RSA at a stationary time $T$ (with absorbing boundary condition in $n=0$), assuming that each species has only one individual when it emerges. Regardless of the type of birth/death rate, such relation can be written as:
\begin{eqnarray}
   \nonumber \mathcal{P}^a_{RSA}(n)&=&\lim_{T\rightarrow+\infty}\frac{1}{\sum_{n\geq1}^{\infty} \int_{0}^T du P(n,T-u|1,0)} \int_{0}^T du P(n,T-u|1,0)=\\
  \label{RSA_stat} &=&\lim_{T\rightarrow+\infty}\frac{1}{f(T)} \int_{0}^T du P(n,u|1,0).
\end{eqnarray}
where $P(k,t|n,t_0)$ is the probability to find $k$ individuals at time $t$ given that there are $n$ at time $t_0<t$. The normalization has been calculated using Eq. (\ref{trick}).

The ST probability function is thus:
\begin{equation}\label{surv_1}
    p_{s}(t)=\sum_{n=1}^{\infty}p_{spt}(t\mid n)\mathcal{P}^{a}_{RSA}(n)=\lim_{T\rightarrow+\infty}\sum_{n=1}^{\infty}\frac{d}{dt}P(0,t|n,0)\lambda\int_0^T du P(n,u|1,0).\quad
\end{equation}

Being at stationarity, time--translational invariance $P(n_2,t_2|n_1,t_1)=P(n_2,t_2-t_1|n_1,0)$ holds:
\begin{eqnarray}
 \nonumber  p_{s}(t) &=&\lim_{T\rightarrow+\infty}\frac{1}{f(T)}\int_0^T\frac{d}{dt} \sum_{n=1}^{\infty}P(0,t|n,0)P(n,0|1,-u)du\\
 \nonumber                &=&\lim_{T\rightarrow+\infty}\frac{1}{f(T)}\int_0^T\frac{d}{dt}\big[P(0,t|1,-u)-P(0,t|0,0)P(0,0|1,-u)\big]du\\
 \nonumber                &=&\lim_{T\rightarrow+\infty}\frac{1}{f(T)}\int_0^T\frac{d}{du}P(0,t+u|1,0)du=\lim_{T\rightarrow+\infty}\frac{1}{f(T)}\int_0^TdP(0,t+u|1,0)=\\
\label{surv_2}            &=&\lim_{T\rightarrow+\infty}\frac{1}{f(T)}\big[P_>(t|1,0)-P_>(t+T|1,0)\big],
\end{eqnarray}
where $P(0,t|1,0)\equiv 1-P_>(t|1,0)$ and $P_>(t|1,0)=\sum_{n\geq1}^{\infty}P(n,t|1,0)=\int_t^{\infty}p_{spt}(\tau)d\tau$.
We thus obtain
\begin{equation}\label{surv_final}
     p_{s}(t)=\frac{1}{\langle \tau \rangle}\int_t^{\infty}p_{spt}(\tau)d\tau,
\end{equation}
and this relation is valid in general, independently of the specific master equations  obeyed by $P(n,t|1,0)$.

\section{Conclusions}
In this paper we have proposed a neutral model for generic birth-death processes, where information on species abundance is subsumed by the SPT distribution, $p_{spt}$, associated to the ecosystem under study. The model can be seen as a coarse grained version of the grand-canonical approach to neutral theory. This framework has two main advantages: 1) Allowed us to obtain analytical results as the transient (and stationary) dynamics of ecosystem species richness, as well as the complete analytical description of the $n-$point correlation function of species diversity; 2) Provide a simple null model, incorporating all the main features of any neutral model based on birth-death processes, as a function of one apriori distribution, namely the SPT distribution. This highlights the important role of $p_{spt}$, as synthetic descriptor of the ecosystem dynamics.  One of the main results obtained is given by the relation eq. (41), valid for any birth and death process, between the persistence time distribution $p_{spt}$ and the survival probability function  $p_{s}$. All the presented results are exact in the assumption that species are noninteracting.  We have compared the analytical results of our model with empirical data on species diversity for a estuarine ecosystem. The analysis shows that, in spite of the minimalist assumptions of our model, complex emergent patterns in the ecosystem dynamics can be captured by the proposed coarse grained neutral framework. This suggest that species diversity data cannot be used  by themselves to discriminate among different type of birth-death processes. Further studies and different approaches are required to determine how species diversity patterns are related to different type of demographic dynamics. Two major simplifications of our analysis are the non-interacting ideal gas like assumption and ignoring the effects of the spatial distribution. Further research will probe what qualitative changes would arise by relaxing the mean field-like  assumption presented here by accounting for dispersal limitation.

\clearpage
\newpage


\clearpage
\newpage

\appendix
\section{}
In this appendix we provide some mathematical details of the results presented in the main text. We start by achieving the probability of having $s$ new species in the ecosystem during the time interval $[0,t)$, $U_s(t)$, given by Eq. (\ref{nu}) in the main text. It's easy to write the ME for $U_s(t)$, $\dot{U}_s(t)=\lambda\big(U_{s-1}(t)-U_{s}(t)\big)$, i.e. new species enter in the system at rate $\lambda$. The corresponding differential equation for the the generating function $\hat{U}(z,t)=\sum_{s=1}^{\infty}z^sU_s(t)$ is $\dot{\hat{U}}(z,t)=\lambda (z-1)\hat{U}(z,t)$ which leads to $\hat{U}(z,t)=e^{\lambda t(z-1)}\hat{U}(z,0)$. Assuming the initial condition $U_s(0)=\delta_K(s-s_0)$, then $\hat{U}(z,0)=z^{s_0}$ and

\begin{equation}\label{gen_nu}
\hat{U}(z,t)=e^{-\lambda t}z^{s_0}e^{\lambda t z}=e^{-\lambda t}\sum_{k=0}^{\infty}\frac{(\lambda t)^k}{k!}z^{s_0+k}=e^{-\lambda t}\sum_{s=s_0}^{\infty}\frac{(\lambda t)^{k-s_0}}{(k-s_0)!}z^{k}.
\end{equation}
From Eq. (\ref{gen_nu}) follows
\begin{equation}\label{nu_2}
    U_k(t|s_0)\left\{
                          \begin{array}{ll}
                            \frac{(\lambda t)^{k-s_0}}{(k-s_0)!}\exp(-\lambda t), & \hbox{for $k\geq s_0$;} \\
                            0, & \hbox{for $k<s_0$,}
                          \end{array}
                        \right.
\end{equation}
which in turn, by noting that $U_k(t)=\sum_{s_0=1}^{\infty}  U_k(t|s_0)U_{s_0}(0)$, leads to Eq. (\ref{nu}). Let's derive now
the probability distribution of the number of persistent species given by Eq. (\ref{p(S,t)1}).

Using Eqs. (\ref{p(S,t)}) and (\ref{nu_2}), the generating function of $\mathcal{P}(s,t)$ reads as:
\begin{eqnarray}\label{gen_fun}
 \nonumber   \hat{ \mathcal{P}}(z,t)&=&\sum_{n=0}^{+\infty} \sum_{s_0=1}^{\infty}\Theta(n-s_0)U_{s_0}(0) e^{-\lambda t}\frac{(\lambda t)^{n-s_0}}{(n-s_0)!}\times\\
&\times& \int_0^t\prod_{i=1}^n \frac{dt_i}{t}\int_0^{\infty}\prod_{j=1}^nd\tau_j p_{spt}(\tau_j)z^{\sum_{i=1}^n\Theta(t_i+\tau_i-t)}.
\end{eqnarray}
Because of the independence of the random variables $t_i$ and $\tau_j$ we can write
\begin{equation}\label{gen_fun1}
    \hat{ \mathcal{P}}(z,t)=\sum_{n=0}^{+\infty} \sum_{s_0=1}^{\infty}\Theta(n-s_0)U_{s_0}(0)\mathcal{I}(z,t)^{s_0} e^{-\lambda t}\frac{(\lambda t)^{n-s_0}}{(n-s_0)!}\mathcal{I}(z,t)^{n-s_0},
\end{equation}
where we have set $t_i=t_0$ and $\tau_i=\tau$ for $i=1,2,...,n$ and $\mathcal{I}(z,t)\equiv
\int_0^t\frac{dt_0}{t}\int_0^{\infty}d\tau p_{spt}(\tau)z^{\Theta(t_0+\tau-t)}$.
Finally, using $\sum_{n=0}^{\infty}\frac{x^n}{n!}=e^x$, and through the relation
\begin{eqnarray}
 \nonumber  z^{\Theta(t_0+\tau-t)} &=& \int_{0}^{t}\Theta(\tau-t)z+\Theta(t-\tau)[z\Theta(t_0-(t-\tau))+\Theta(t-(t_0+\tau))]dt_0\qquad\\
\label{ztheta1}&=&zt\Theta(\tau-t)+(1-\Theta(\tau-t))(z\tau+t-\tau)
\end{eqnarray}
we have
\begin{eqnarray}
 \nonumber \mathcal{I}(z,t)&=&\int_0^t\frac{dt_0}{t}\int_0^{\infty}d\tau p_{spt}(\tau)zt\Theta(\tau-t)+(1-\Theta(\tau-t))(z\tau+t-\tau)\\
\label{I} &=&(z-1)\left(\int_0^t d\tau\tau p_{spt}(\tau)+t(P_>(t)-1)\right)=(z-1)f(t)-t,
\end{eqnarray}
where we have used $\int_0^t d\tau\tau p_{spt}(\tau)=-\int_{0}^{t}d\tau\tau \dot{P}_{>}(\tau)=-t P_{>}(t)+\int_{0}^{t}d\tau P_{>}(\tau)$ and
\begin{equation}\label{f(t)A}
   f(t)\equiv\int_{0}^{t}d\tau P_{>}(\tau)=\int_0^{+\infty}p_{spt}(\tau)\min[t,\tau]d\tau,
\end{equation}
that is Eq. (\ref{f(t)}) in the main text.

Substituting Eqs. (\ref{f(t)A}) and (\ref{I}) in  Eq. (\ref{gen_fun1}), together with the observation that
\begin{equation}\label{f(t)A}
  \sum_{s_0=1}^{\infty}\Theta(n-s_0)U_{s_0}(0)\mathcal{I}(z,t)^{s_0} =   \hat{U}(1+\frac{z-1}{t}f(t),0)
\end{equation}
leads to Eq. (\ref{log_gen_fun4}) presented in the main text.

\newpage
\clearpage
\section*{References}
\bibliographystyle{jphysicsB}

\end{document}